\documentstyle[prd,aps,preprint]{revtex}
\begin{document}
\draft
\title{\bf A Maxwellian Path to the
$q$-Nonextensive Velocity Distribution
Function }

\author{R. Silva $^{1}$\footnote[1]
{raimundo@dfte.ufrn.br} A. R.
Plastino$^{2}$\footnote[2]
{plastino@fcaglp.fcaglp.unlp.edu.ar}
and J. A. S. Lima$^{1}$\footnote[3]
{limajas@dfte.ufrn.br}}

\smallskip
\address{~\\$^1$Universidade
Federal do
Rio Grande do Norte,
\\Departamento de F\'{\i}sica,
Caixa Postal 1641, \\59072-970 Natal, RN,
Brazil}
\address{~\\$^2$Facultad de Ciencias
Astronomicas y Geofisicas
\\Universidad Nacional de La Plata,
C.C. 727, 1900 La Plata, Argentina}

\date{\today}
\maketitle

\vskip 1.5cm
\begin{abstract}

Maxwell's first derivation of the
equilibrium distribution function for a
dilute gas is generalized in the spirit
of the nonextensive
$q$-statistics proposed by Tsallis. As
an application, the $q$-Doppler
broadening of spectral lines due to
the random thermal motion of the
radiating atoms is derived.

\end{abstract}


It is widely known that the
thermodynamical or statistical
description
of nonextensive systems demand
a generalization of the usual
Boltzmann-Gibbs
thermostatistics \cite{tsa1,tsa2,tsa3}.
A few important examples of physical
systems or processes
where the standard approach seems
to be inadequate are self-gravitating
systems, some kinds of plasma
turbulence, and self-organized criticality.

The standard Boltzmann-Gibbs
approach is based on the extensive entropy
measure

\begin{equation}\label{eq:S}
S = -k\sum_i p_i\ln{p_i} \quad,
\end{equation}
where $k$ is the Boltzmann constant
and $\{p_i\}$ denotes the
probabilities of the microscopic
configurations. Ten years ago, in order
to deal with the above mentioned
difficulties associated with
nonextensivity, Tsallis\cite{tsallis,tsacu}
proposed the following
nonextensive form of entropy
\begin{equation}\label {eq:Sq}
S_q = k\frac{\left[1-\sum_i p_i^{q}\right]}{(q-1)} \quad,
\end{equation}
where $q$ is a parameter
quantifying the degree of
nonextensivity. For
instance, given a composite
system $A+B$, constituted by
two subsystems
$A$ and $B$, which are independent
in the sense of factorizability of
the microstate probabilities
(i.e. $P_{ij}^{(A+B)}=P^{(A)}_i P^{(B)}_j
$), the Tsallis measure verifies

\begin{equation}
S_q(A + B) = S_{q}(A) +S_{q}(B) + (1-q)S_q(A)S_q(B)
\end{equation}

In the limit $q\rightarrow 1$, $S_q$ reduces
to the standard logarithmic
measure (1), and the usual additivity of
entropy is recovered.
There is a growing body of evidence
suggesting that the $q$-entropy
may provide
a convenient frame for the
thermostatistical analysis of many
 physical scenarios, such as stellar polytropes \cite{plastino},
turbulence in electronic
plasmas \cite{turbulence}, anomalous
diffusion
\cite{walks}, Levy
distributions \cite{levy,c95}, the
critical regime in low
dimensional dissipative chaotic systems \cite{zheng,chaos,l98},
the solar neutrino problem \cite{klq96},
peculiar velocity distribution of
galaxy clusters \cite{lk98} or
more generally, systems endowed with long range interactions,
long range memory effects, or a
fractal-like space-time
\cite{tsa1,tsa2}.  The nonextensive
thermostatistics has
been shown to be endowed with
interesting mathematical properties
\cite{ragg1,ragg2,ragg3}, the
main theorems of the standard
statistics admiting suitable
generalizations
\cite{chame,rajago,r96,l97}.
The issue related with the
connection between $q$-statistics and
$q$-thermodynamics has also been
addressed\cite{tsacu,PP77}, while the
time
evolution of $S_q$ has been
analyzed both in the discrete
case through a
direct application of the master
equation\cite{mariz}, and in the
continuous one in connection with
the Liouville and Fokker-Planck
equations \cite{bpt98}. Some
cosmological implications of Tsallis
generalized thermostatistics
have also been discussed \cite{h96,t97}.
However, at present there is only
a limited understanding on the
relation between  the $q$ parameter
and the underlying microscopic
dynamics. In the cases of low
dimensional dissipative chaotic maps
\cite{zheng,chaos}, and in some
toy models of self-organized
criticallity \cite{pajarito}, the
value of $q$ characterizing the
 system has been
obtained from studies of the
concomitant dynamics.
In spite of the importance of
these developments, they do not involve
directly the Tsallis maximum entropy
distribution and the experimental
evidences supporting it.
In order to clarify this point, let
us briefly review the main
observational facts supporting
nowadays Tsallis'
proposal. First, a  Tsallis'
maximum entropy distribution
has been shown
to describe properly
a metastable equilibrium state
of a 2-dimensional pure electron
plasma \cite{turbulence}.
Second, the  $q$-distribution
corresponding to
an ideal classical gas
provides a better fit for the observed
distribution of peculiar
velocities of galaxy clusters
than the ones
obtained by recourse to more complicated
models based on the standard
thermostatistics \cite{lk98}.
Finally, assuming a $q$-velocity
distribution for the involved
particles, the evaluation of the nuclear
reaction rates in the solar
interior predicts a neutrino flux in
agreement with the observational
data, thereby suggesting that
Tsallis' thermostatistics may provide a
solution for the well-known solar neutrino problem\cite{klq96}.

It is remarkable that all the
experimental evidence listed above deals,
directly or indirectly, with
the $q$-distribution of velocities, which can
be obtained maximizing $S_q$
under the normalization
and mean  energy
constraints. Within a more general framework,
such a distribution
describes how the $q$-nonextensive
canonical ensemble,
associated with the classical
many body problem, depends on
the particle
velocities \cite{ppt94}. In
this way, the $q$-velocity distribution seems
to be a reasonable  nonextensive
generalization of the celebrated
Maxwell-Boltzmann distribution, which
is recovered as the particular $q
\rightarrow 1$ limiting case.
In spite of its theoretical
interest, a satisfactory microscopic
explanation for the physical
origin of the $q$-velocity distribution is
still lacking, although some
interesting attempts have recently been
made in connection with the linear
and nonlinear Fokker-Planck
equations \cite{b98}. However, to
shed some light on this matters, it
seems important to consider
suitable (nonextensive) generalizations of
the kinetic approach pioneered by
Maxwell and Boltzmann.

In this letter, we are interested in
exploring the kinetic route. Our
aim is to rediscuss the correspondence
between the parameter $q$
introduced by Tsallis and the $q$-equilibrium
velocity distribution for a
Maxwellian gas, however, assuming
from the very beginning a nonextensive
generalization of the
separability hyphothesis
originally proposed by
Maxwell\cite{M1}. Hopefully, as
happened in the extensive framework,
this line of inquiry may
provide some insight for a more rigorous
kinetic irreversible treatment
from the Boltzmann viewpoint. As a new
application, we deduce a formula for the $q$-Doppler broadening of
spectral lines.

Let us now consider a spatially
homogeneous gas, suposed in equilibrium
at temperature $T$, in such a
way that $F({\bf v})d^{3}{\it v}$ is the
number of particles with
velocity in the volume element $d^{3}{\it v}$
around ${\bf v}$. In Maxwell's
derivation, the 3-dimensional
distribution is factorized
(lottery assumption) and depends only on the
magnitude of the velocity\cite{M1,Sommerf}
\begin{equation}\label{eq:F}
F \left( \sqrt{v_x^2+v_y^2 + v_z^2 } \right)
d^3 v =f(v_x)f(v_y)f(v_z)dv_xdv_ydv_z
\quad,
\end{equation}
from which it is straightforward to show that
\begin{equation}\label{eq:F1}
f(v_i) = A_1 e^{-{\beta mv_i^2 \over 2}}\quad,
\quad i= x,y,z
\end{equation}
and
\begin{equation}\label{eq:F2}
F({\bf v}) = A_1^{3}
e^{-{\beta mv^2 \over 2}} \quad,
\end{equation}
where $\beta = {1 \over kT}$ in order
to recover the standard
macroscopic thermodynamic
relations, and
$A_1=\sqrt {m \over 2\pi kT}$
is the normalization constant.
Naturally, in the nonextensive context
described by (2), the starting
basic hypothesis (4), which takes
into account the isotropy of all
velocity directions, must somewhat
be modified.
Physically, Maxwell's ansatz is tantamount
to assume that the three components of
the velocity are uncorrelated. However,
this property does not hold in the systems
endowed with long range interactions where
Tsallis distribution has been observed
\cite{turbulence,klq96,lk98}.
Notice that the Maxwell ansatz is
equivalent to  express
$ \ln F $ as the sum of the
logarithms of the one dimensional
distribution functions associated
with each velocity component.
A simple and natural way to generalize
this procedure within the nonextensive
formalism, in order to introduce
correlations between the velocity
components, would be to replace the
logarithm function by a power law.
  However, in order to recover the
ordinary logarithmic ansatz as a
particular limiting case, it is
convenient to express the power
generalization in terms of the $q$-log
function, which is a combination
of a power function plus appropriate
constants.
Elementary considerations may
convince oneself that a consistent
$q$-generalization of (4) is
(for simplicity we provisionally consider
the bidimensional case)
\begin{equation} \label{eq:2.7}
F\left( \sqrt{v_x^2+v_y^2} \right) d^2v =
e_q(f^{q-1}(v_x)\ln_q f(v_x)+f^{q-1}(v_y)\ln_q
f(v_y))dv_x dv_y \quad,
\end{equation}
where the $q$-exp and $q$-log functions, $e_q(f)$, $\ln_q(f)$, are
defined by
\begin{equation} \label{eq:1}
e_q(f)=[1+(1-q)f]^{1/1- q}\quad,
\end{equation}
\begin{equation} \label{eq:2}
\ln_q f={f^{1-q}-1\over 1-q}\quad.
\end{equation}
As one may check,
$e_q(\ln_q f) = \ln_q (e_q(f)) = f$, and from (9) we
see  that the $q$-log
differentiation,
${d \over dx}\ln_q f =
f^{-q}{df \over dx}$ is also
satisfied. Note also
that in the limit ${q \rightarrow 1}$
the identities (8)-(9) reproduce
the usual properties of the
exponential and logarithm functions, and
(7) reduces to the bidimensional
case of (4), as should be expected.
Now, partial $q$-log differentiation
of (7) with respect to $v_x$ yields

\begin{equation} \label{eq:2.11}
{\partial \ln_q F\over\partial v_x}={\partial
\ln_q[e_q(f^{q-1}(v_x)\ln_q f(v_x)+f^{q-1}(v_y)\ln_q
f(v_y))]\over\partial v_x} \quad,
\end{equation}
or equivalently,
\begin{equation} \label{eq:2.12}
{v_x\over \chi}{F'(\chi)\over F^q(\chi)}={\partial\over\partial
v_x}\{f^{q-1}(v_x)\ln_{q}f(v_x)\}\quad,
\end{equation}
where $\chi = \sqrt{v_x^2+v_y^2}$ and
a prime means total derivative.
Note that analogous equations
apply to the remaining components
even whether we had considered the
n-dimensional case. Introducing the
shorthand notation
\begin{equation} \label{eq:16}
\Phi(\chi)={1\over \chi}
{F'(\chi)\over F^q(\chi)}\quad,
\end{equation}
we may rewrite (11) as

\begin{equation} \label{eq:2.17}
\Phi(\chi)={1 \over v_x}{\partial\over\partial
v_x}\{f^{q-1}(v_x)\ln_{q}f(v_x)\}
={1 \over v_y}{\partial\over\partial
v_y}\{f^{q-1}(v_y)\ln_{q}f(v_y)\}.
\end{equation}

\noindent
The second member of the above
equation only depends on $v_x $, while
the
third one is a function exclusively
of variable $v_y $. Hence, equation
(\ref{eq:2.17}) can be satisfied
only if all its members are equal to
one
and the same constant, not depending on
any of the velocity components.
So,
we may put $\Phi (\chi)=-m\gamma$, where
$m$ is the mass of the particles
and $\gamma$ is an arbitrary constant. Of
course, the introduction of
$m$
at this point is dictated only by
the known Maxwellian limit. As one may
see from (9),
$f^{q-1}(v_x)\ln_{q}f(v_x) =
\ln_{q^*} f(v_x)$, where $q^{*}=2-q$.
Hence, the general solutions for
$f(v_x)$ is given by (equivalent
expressions are valid for $f(v_y)$
and $f(v_z)$)
\begin{equation} \label{eq:2.21}
\ln_{q^*}f(v_x) = -\frac{m\gamma}{2} v_x^2+\ln_{q^*} A \quad,
\end{equation}
where, without loss of generality, we
have written the integration
constant in a convenient form. Now, taking
the q-exponential in both
sides of (14) it follows that

\begin{equation} \label{eq: 2.23}
f(v_x)=\left[1+(1-q^*)\left(\ln_{q^*} A-\frac{\gamma
mv_x^2}{2}\right)\right]^{1/1-q^*}\quad,
\end{equation}
and defining a new constant $\beta$ as
\begin{equation} \label{eq: 2.25}
\beta={\gamma\over 1+(1-q^*)\ln_{q^*} A}={\gamma\over A^{1-q^*}}\quad,
\end{equation}
we find the generalized expression
\begin{equation} \label{eq: 2.26}
f(v_x)= A_q\left[1 - (q-1)\frac
{\beta mv_x^2}{2}\right]^{1 \over {q-1}}
\quad,
\end{equation}
where we have introduced a subindex $q$
to make explicit the
$q$-dependence of $A$. From (17) we see
that the Gaussian probability
curve of the Maxwellian gas is replaced by
the charactheristic power law behavior of Tsallis' nonextensive
framework,
and as expected, the limit $q=1$
recovers the exponential extensive
result. Note also that for values
of $q$ greater than unity, the
positiviness of power argument means
that (17) exhibits a thermal
cut-off in the maximal allowed
velocities. The components of the
velocities lie on the interval
$[-L, L]$, where
$L =\sqrt{2 \over m\beta(q-1)}$.
Hence, the integration
limits in the standard normalization
condition is modified in such a way
that only if $q=1$ they go to infinity. Taking this into account one may
show that the normalization constant $A_q$ can be expressed in terms of
Gamma-functions as
\begin{equation}
{A_q} = \left( {{1+q}\over 2} \right)
{\Gamma({1\over 2}+{1\over {q-1}})\over \Gamma({1\over {q-1}})}
\sqrt{m(q-1)\over 2\pi kT}  \quad.
\end{equation}
Further, using that $\lim_{|z|\rightarrow \infty} {\Gamma(a+z)\over
\Gamma(z)}e^{-a\ln z}=1$ (see Abramowitz\cite{Abramowitz}), it is easy
to see that $A_1$ is the standard Maxwellian result.

Before continuing we need to
obtain the complete distribution. By adding
the $v_z$ component to (7) it is
readily seen that
\begin{equation}  \label{eq: 2.26c}
F \left( \sqrt{v_x^2+v_y^2+v_z^2} \right) d^2v =
e_{q}(\ln_{q^*} f(v_x) + \ln_{q^*}
f(v_y)+\ln_{q^*} f(v_z))dv_x dv_y dv_z\quad.
\end{equation}
Hence, taking the q-logarithim of
the above expression and repeating the
same algebraic steps of the
one-dimensional case it follows that
\begin{equation} \label{eq: 2.33}
F({\bf v})=B_q \left[1 -
(q-1)\frac{\beta mv^2}{2} \right]^{1 \over {q-1}}
\quad,
\end{equation}
where $B_q$ is fixed by the 3-dimensional
normalization condition. We
find
\begin{equation} \label{bq}
B_q = (q-1)^{1/2} \, {\frac{(3q-1)}{2}}
\left( {{1+q}\over 2} \right)
{\Gamma({1\over 2}+{1\over {q-1}})\over
\Gamma({1\over {q-1}})}
\left[\frac{m}{2\pi kT} \right]^{\frac{3}{2}}
\quad.
\end{equation}
As expected, the q-distribution (20) is
isotropic meaning that all
velocity
directions are also equivalent in this
generalized context.
As remarked earlier for the
one-dimensional case, there also exist a
temperature dependent cut-off on
the magnitude of the
velocities. From (18)
we see that the $B_1=A_1^{3}=[\frac{m}{2 \pi kT}]^{\frac{3}{2}}$ is
the standard 3-dimensional
Maxwellian result as it should be.

Now we discuss an important point
of principle, which is related with
the definition of marginal
probabilities in the context of
$q$-statistics. In ordinary
space, a marginal probability, say,
$f(v_x)$, may be obtained
from the $3$-dimensional distribution by
\begin{equation}
f(v_x) = \int F(v_x, v_y, v_z)dv_y dv_z \quad.
\end{equation}
This elementary and natural definition,
widely applied in statistical physics,
is not usually regarded as deserving
any further scrutiny. However,
within the nonextensive framework
we are discussing here,
the concept of marginal probability
distributions shows some new
remarkable features.
Basically, this occurs because the
distribution $f(v_x)$, as given
by (22), does not coincid with our
equation (17). As one may check, it
has a power different from ${1 \over
{q-1}}$. In other words, using the
above formula, the 3-dimensional
distribution (20) leads to a $f(v_x)$
with a different value of $q$.
Strictly speaking, since the
q-distribution cannot be factorized, the
power of the one-dimensional
distribution obtained from a marginal
probability like (22), depends
on the number of spatial dimensions.
The power increases by ${1 \over 2}$
for each additional dimension
present in the complete distribution. This
is equivalent to say that
$f(v_x)$ is also a power law, but
with an effective $q$-parameter.
However, to interpret this fact in
a consistent way, it is required to
introduce an effective temperature
in the one-dimensional distribution.
Naturally, the same happens when we
consider arbitrary dimensions $m$,
$n$, where $m<n$.  In particular, this
means that the zeroth law of
thermodynamics is not satisfied for
systems described in this
nonextensive framework. Hopefully, a
proper modification of (22) can be
found which avoids these undesirable
features, but preserves the
interesting ones. In this concern, we
suggest the following general
expression for the marginal $m$-dimensional
distribution in a
$n$-dimensional $q$-velocity space
\begin{equation}
f(v_1, v_2,...,v_m) =
\frac{\int F^{\alpha}(v_1, v_2,...,
v_n)dv_{m+1}dv_{m+2}
\ldots dv_n}{\int F^{\alpha}(v_1, v_2,...,
v_n)dv_{1}dv_{2}...dv_n} \quad,
\end{equation}
where $\alpha=1 - {1 \over 2}(q-1)(n-m)$.
For $q=1$, we have $\alpha=1$,
and the standard definition is recovered.
In the above discussed case,
$n=3$,
$m=1$, one finds $\alpha=2-q$. For
this value of $\alpha$ it is
straightforward to show that (23)
reproduces the one-dimensional
distribution (17). As a matter
of fact, our prescription (23)
solves the
conflict with the zeroth law of
thermodynamics, in the sense that the
same power law and temperature of
the complete distribution is always
obtained regardless of the specific
dimensions involved in the problem.
Besides, it is interesting to realize
that equation (23) can be
interpreted as defining the ordinary
marginal probability function
computed using the escort
distribution \cite{bs95}

\begin{equation}
F^* \, = \, \frac{F^\alpha }
{\int F^\alpha d^N v},
\end{equation}

\noindent
instead of being evaluated using
the original distribution $F$. This is
a standard procedure in the fractal
thermodynamic formalism \cite{bs95}.

\noindent {\bf Broadening of Spectral Lines}.
The random motion of
particles broadens spectral lines, first,
because of collisions between
the particles (pressure broadening), and
second, due to the thermal
Doppler effect of the radiating atoms. As
widely known, in the extensive
case the first effect is proportional to
$pT^{-{1 \over 2}}$, where p is
the pressure, whereas the second scales
with $T^{1 \over 2}$. Let us now
discuss the latter effect using the
$q$-Maxwellian velocity distribution.
The standard result was
derived by Lord Rayleigh and further
observed by Michelson\cite{Ray,Mike}.

In order to estimate the magnitude of
the $q$-Doppler broadening for the
visible light emitted by the molecules
of a hot
gas, it is enough to consider the one
dimensional
case. Neglecting relativistic effects, the
frequency shift viewed along
the $x$ direction is given by the standard
Newtonian formula
\begin{equation} \label{eq: 2.35}
\nu=\nu_0\left(1+{v_x\over c}\right)\quad.
\end{equation}
The frequency distribution expected for a
spectral line centered at
$\nu_0$ is obtained changing variables
from $v_x$ to $\nu$. From (17) it
is readily seen that
\begin{equation} \label{eq: 2.38}
f(\nu)d\nu=A_q\left[1 - (q-1){mc^2\over
2kT}\left({\nu-\nu_o\over\nu_o}
\right)^2\right]^{1\over
q-1}{c\over\nu_o}d\nu\quad,
\end{equation}
which has also the form of a Maxwellian
$q$-distribution. The broadening is usually
measured by the width of the
spectral line at half intensity. It is easy
to check that in terms of the
wavelenght, the standard deviation is
replaced by

\begin{equation} \label{eq: 2.39}
\Delta\lambda_d=\lambda_o
\left[{2kT\over mc^2}\left({1-2^{1-q} \over
q-1}\right)\right]^{1/2}=
\lambda_o\left({kT\over mc^2}2 \ln_{q}
2\right)^{1/2}\quad.
\end{equation}
For $q=1$ the standard result is
recovered as should be
expected\cite{Sommerf,Ray}. However, although
mantaining the same $T^{1
\over 2}$ temperature dependence, the
thermal Doppler broadening is
modified in the nonextensive
framework. Naturally, the above formula can
be used to limit the q-parameter. Note
that in a log-log plot of
$\Delta\lambda_d$ versus ${kT \over mc^2}$,
the straight line is
displaced  parallel to itself for each
value of $q$ different from
unity. In comparison with the standard
$q=1$ result, the above
$q$-velocity distribution gives a
narrowing of the Doppler width for
$q<1$, and a broadening tendency for
$q$ larger than unity.

\noindent {\bf Tsallis' Generalized
Mean Values.} It is worth noticing
that we might have considered a different
generalization for the factorization
condition. In principle, instead of
equation
(\ref{eq:2.7}) one may assume the
following ansatz

\begin{equation}  \label{qqfact}
F \, = \, e_q( \ln_q f(v_x) + \ln_q f(v_y) ).
\end{equation}

\noindent
In this case, repeating the same steps
we have already explained,
instead of (\ref{eq: 2.26}), one obtains
a slightly different velocity
distribution

\begin{equation} \label{qqdist}
f(v_x)= A_q \left[  1- (1-q) \, \frac{1}{2}
\, \beta \,
mv_x^ 2   \right]^{\frac{1}{1-q}},
\end{equation}

\noindent
which is exactly that one determined by
Tsallis MaxEnt
prescription when the generalized
mean value\cite{tsacu}

\begin{equation} \label{qqmean}
\langle v_x^2 \rangle_q \, = \,
\int \, f^q \, v_x^2 \, dv_x
\end{equation}

\noindent
is a meaningful constraint. On
the other hand, by employing the standard
linear mean values, distribution
(\ref{eq: 2.26}) is obtained.
As a matter of fact, none of the
main results and conclusions of this
paper change in
a significative way whether one
adopts the alternative factorization
prescription
(\ref{qqfact}). For instance, all
the results
corresponding to our analysis of
the $q$-Doppler broadening follow
simply by replacing everywhere
$(q-1)$ by $(1-q)$. This is a strong
indication that a more conclusive
result needs a full kinetic theoretic
treatment, which requires a proper
generalization of the Boltzmann
$H$-theorem.

It is possible that the approach
developed here may be implemented
even if more general expressions
for nonextensive entropies, which
reduce in a commom limit to the
Boltzmann-Gibbs-Shanon form,
are considered. Tsallis measure $S_q$
is not the only conceivable mathematical
generalization of the
standard logarithmic entropy.
Other interesting nonextensive entropic
functionals have been recently
proposed \cite{A97,P98}. However,
Tsallis entropy has been shown to
be endowed with many elegant and useful
mathematical properties, and its
associated $q$-MaxEnt
distributions have been experimentally
observed \cite{turbulence,lk98}. It would
be of great interest to explore the,
so far poorly known,
mathematical properties of the
recently introduced
entropies \cite{A97,P98}, as
well as to determine
if they admit relevant
physical applications.

\noindent {\bf Conclusions.} In the
present work we have obtained
Tsallis non-extensive velocity
distribution by recourse to an
argument akin to the celebrated
derivation
advanced by Maxwell for the
equilibrium velocity
distribution. As shown
by  Maxwell, his distribution is
the only one compatible with
isotropy and factorizability
with respect to
each velocity component.
Similarly, Tsallis
q-distribution is uniquely
determined by the requirements of (i)
isotropy,
and (ii) a suitable generalization
of the factorizability condition.
Maxwell's factorization condition is
tantamount in requiring that the
{\it
logarithm} of the complete distribution
function be equal to a sum of
$N$
terms, each one depending only on one
velocity component. Instead of
logarithm,
our factorization condition requires that
a {\it power} of the complete
distribution be equal to a sum of $N$
terms, which also depend only
on one velocity component. Reformulating
this
last condition in terms  of the
Tsallis $q$-logarithm function,
Maxwell's
expressions are recovered in the
$q\rightarrow 1$ limit. The same happens
with the formula giving the
broadening of spectral lines.

It is important to stress  that
the simple transformation
$(q-1) \, \rightarrow \, (1-q)$ is enough
to recast all the present results within the
complete Tsallis formalism \cite{tsacu}
based on Tsallis generalized mean
values. In that case, Tsallis
$q$-distribution would adopt the form

\begin{equation} \label{qvevq}
F({\bf v})=\tilde B_q \left[1 -
(1-q)\frac{\beta mv^2}{2} \right]^{1 \over {1-q}}
\quad,
\end{equation}

\noindent
where $\tilde B_q $ is given by the expression
obtained from (\ref{bq}) after replacing $(q-1)$
by $(1-q)$. In a similar vein, the $q$-Doppler
width of spectral lines would appear under the
guise

\begin{equation} \label{qiliq}
\Delta\lambda_d=
\lambda_o\left({kT\over mc^2}\, 2^q\, \ln_{q}
2\right)^{1/2}\quad.
\end{equation}

  Mathematically, Maxwell factorization
condition is similar to the
Maxwell-Boltzmann hypothesis
 of ``molecular chaos", which
assumes that the
two-molecule distribution function
describing the colliding molecules
is factorizable as the product of
two one-molecule distributions
(i.e. $F({\bf v}_1,{\bf v}_2)  =
f({\bf v}_1) \, f({\bf v}_2) $).
This hypothesis plays a fundamental
role in the standard kinetic theory of
gases \cite{Sommerf}.
Boltzmann' proof that any initial
velocity distribution evolves
irreversibly towards Maxwell's
distribution does not rely
just in the general principles
of classical mechanics. It also
needs the additional assumption
of ``molecular chaos". Boltzmann
himself \cite{B64} recognized that the
hypothesis of ``molecular chaos"
may not always hold, especially at
high densities.
  Our present results suggest that Boltzmann's approach to
Maxwell's velocity distribution
can be adapted to the non-extensive
setting by recourse to an appropriate
generalization of the ``molecular chaos"
assumption. This issue will be addressed in a forthcoming communication.

\noindent
{\bf Acknowledgments:} It is a
pleasure to thank Constantino Tsallis and
Janilo Santos for valuable
discussions. This work was partially
supported by
Pronex/FINEP (No. 41.96.0908.00),
Conselho Nacional de
Desenvolvimento Cient\'{i}fico e
Tecnol\'{o}gico - CNPq and
CAPES (Brazilian Research Agencies), and by
CONICET (Argentina Research Agency).

\end{document}